\documentclass[10pt]{emulateapj}
\usepackage{apjfonts}
\usepackage{psfig}
\usepackage{epsfig}

\newcommand\puncspace{\ifmmode\,\else{\ifcat.\C{\if.\C\else\if,\C\else\if?\C\else%
\if:\C\else\if;\C\else\if-\C\else\if)\C\else\if/\C\else\if]\C\else\if'\C%
\else\space\fi\fi\fi\fi\fi\fi\fi\fi\fi\fi}%
\else\if\empty\C\else\if\space\C\else\space\fi\fi\fi}\fi}
\newcommand\SP{\let\\=\empty\futurelet\C\puncspace}
\font\smfont=cmr9
\newcommand\Halpha{{H$\alpha$}\SP}                     
\newcommand\I{\kern.2em{\smfont I}\SP}                 
\newcommand\II{\kern.2em{\smfont II}\SP}               
\newcommand\fnii{\mbox{\rm [N\II]}\SP}                 
\newcommand\foii{\mbox{\rm [O\II]}\SP}                 
\newcommand\HI{\mbox{\rm H\I}\SP}                      
\newcommand\HII{\mbox{\rm H\II}\SP}                    
\newcommand\Ib  {$I$-band\SP}                          
\newcommand\Bb  {$B$-band\SP}                          
\newcommand\BI  {$B-I$\SP}                             
\newcommand\cf{\textit{cf.} }                          
\newcommand\eg{\textit{e.g.}, }                        
\newcommand\etal{\textit{et al.} }                     
\newcommand\ie{\textit{i.e.} }                         
\newcommand\pone  {Paper I}                            
\newcommand\ptwo  {Paper II}                           
\newcommand\UGC {UGC\kern.33em}                        
\newcommand\kms {km~s$^{-1}$\SP}                       
\newcommand\hMpc{h$^{-1}$~Mpc\SP}                      
\newcommand\hkpc{h$^{-1}$~kpc\SP}                      
\newcommand\dV  {$\Delta V_{circ}$\SP}                 
\newcommand\Hax {H$\alpha_{ext}$\SP}                   
\newcommand\aas {A\&AS}                                
\newcommand\aspcs {in ASP Conf. Ser. }                 

\shorttitle{Properties of Cluster Spirals III.}
\shortauthors{Vogt et al.}

\slugcomment{Accepted by AJ 2004 February 23}

\begin{document} 

\title{M/L, \Halpha Rotation Curves, and \HI Gas Measurements     
       for 329 Nearby Cluster and Field Spirals:                  
       III. Evolution in Fundamental Galaxy Parameters}

\author{Nicole P. Vogt\altaffilmark{1,2}}
\affil{Department of Astronomy, New Mexico State University, Las Cruces, NM 88003}
\email{nicole@nmsu.edu}
\and
\author{Martha P. Haynes,\altaffilmark{3} Riccardo Giovanelli,\altaffilmark{3}
and Terry Herter}
\affil{Center for Radiophysics and Space Research, Cornell University, 
Ithaca, NY 14853}
\email{haynes, riccardo, and herter@astro.cornell.edu}

\altaffiltext{1}{Formerly at: Institute of Astronomy, University of Cambridge, Cambridge, 
CB3$-$0HA, UK}
\altaffiltext{2}{Formerly at: Center for Radiophysics and Space Research, 
Cornell University, Ithaca, NY 14853}
\altaffiltext{3}{National Astronomy and Ionosphere Center;  NAIC is operated 
by Cornell University under a cooperative agreement with the National Science 
Foundation.}

\begin{abstract}
We have conducted a study of optical and \HI properties of spiral galaxies
(size, luminosity, \Halpha flux distribution, circular velocity, \HI gas mass)
to investigate causes (\eg nature versus nurture) for variation within the
cluster environment.
We find \HI deficient cluster galaxies to be offset in Fundamental Plane
space, with disk scale lengths decreased by a factor of $25$\%.  This may be a
relic of early galaxy formation, caused by the disk coalescing out of a
smaller, denser halo (\eg higher concentration index) or by truncation of the
hot gas envelope due to the enhanced local density of neighbors, though we
cannot completely rule out the effect of the gas stripping process.  The
spatial extent of \Halpha flux and the \Bb radius also decreases, but only in
early type spirals, suggesting that gas removal is less efficient within
steeper potential wells (or that stripped late type spirals are quickly
rendered unrecognizable).  We find no significant trend in stellar
mass-to-light ratios or circular velocities with \HI gas content,
morphological type, or clustercentric radius, for star forming spiral galaxies
throughout the clusters.
These data support the findings of a companion paper that gas stripping
promotes a rapid truncation of star formation across the disk, and could be
interpreted as weak support for dark matter domination over baryons in the
inner regions of spiral galaxies.
\end{abstract}


\renewcommand{\floatpagefraction}{0.70}
\section{Introduction}

Recent years have seen the emergence of a standard model for the growth of
structure -- the hierarchical clustering model -- in which the gravitational
effects of dark matter drive the evolution of structure from the near-uniform
recombination epoch until the present day.  Simple models for galaxy formation
in the context of these CDM cosmogonies have been remarkably successful in
reproducing the properties of the local galaxy population (Kauffmann, White,
\& Guiderdoni 1993; Cole \etal 1994) and have been extended to predict the
sizes, surface densities and circular velocities of spiral galaxy disks
(Dalcanton, Spergel, \& Summers 1997; Mo, Mao, \& White 1998).

The result has been a testable scenario that predicts the basic structural
properties of the disk galaxy population in any specific cosmogony of CDM
type, and how the ensemble of disk galaxies should evolve with redshift.  The
models assume that most spirals formed as the central galaxies of isolated
halos, an assumption supported by the fact that they must have undergone
minimal dynamical disturbance since the formation of the bulk of the disk
stars (T\'oth \& Ostriker 1992).  The size of the disk is thus expected to
scale with that of its halo, and at high redshift the predicted distribution
of halo sizes is shifted to smaller values.

By $z$ = 1 the predicted change in disk size is almost a factor of 3 for an
Einstein-de Sitter Universe, though only about 1.5 for a flat universe with
$\Omega_0$ = 0.3.  High redshift field spirals show some evidence for this
effect (Vogt 2000), though surface brightness selection biases can mimic this
trend and it is difficult to separate the two factors (Vogt \etal 2004c).
Alternatively, one can search for evidence within the fossil record of local
clusters.  Specifically, spiral galaxies which formed early in the vicinity of
rich clusters may preserve signatures of early coalescence (Kauffmann 1995),
even as we observe them today.  The difficulty, of course, is to distinguish
between such cosmological variations in disk structure and in the direct
effects of the cluster environment.  Some fundamental disk properties (\eg
scale length) may prove to be more a function of the cluster-wide environment
than of the galaxy-specific environment, while the properties of the gas
reservoir are strongly dependent upon the individual merger history and on
interactions with the intracluster medium.

There have been a number of theoretical approaches to modeling the
Tully-Fisher (Tully \& Fisher 1977) relation and its implications for disk
formation and evolution (Rhee 1996; McGaugh 2000; van den Bosch 2000 and
references therein).  Its successful use to probe the peculiar velocity field
about clusters is dependent on the assumption that the relevant galaxy
properties do not vary significantly in different environments.  Ideally
Tully-Fisher studies utilize many galaxies within a cluster to reduce the
distance error estimate; by measuring velocity widths from \Halpha rotation
curves, \HI selected samples can be augmented with gas-poor galaxies for which
the \HI line profile cannot furnish a velocity width (\cf Giovanelli \etal
1997a).  This increases the sampling of a cluster, and in addition \HI
stripped spirals are the most likely to be true cluster members rather than
members of infalling groups, offset both on the sky and in projection.

Evidence for systematic changes in the Tully-Fisher relation as a function of
environment could invalidate such studies.  Variations in the mass and light
distribution of galaxies with environment could thus have a significant
effect, as the Tully-Fisher relation might vary not only between the field and
clusters, but, if environmental effects are important, from cluster to cluster
depending upon its evolutionary state.  Furthermore, Salucci, Frenck, \&
Persic (1993) argue that the scatter in the Tully-Fisher relation can be
decreased by decomposing the mass distribution within galaxies into a dark and
a luminous component to isolate the contribution of the disk alone to the
circular velocity of the system (see however Jablonka \& Arimoto 1992),
implicitly emphasizing the need for detailed knowledge of the environmental
variations of the dark and luminous mass distributions.  Though most of the
Tully-Fisher studies performed to date have not contained large or
sufficiently well--defined samples to properly assess environmental impacts,
there is some evidence (Biviano \etal 1990) against a strong dependence of the
Tully-Fisher relation calibration upon mean cluster densities, radial
positions of galaxies within clusters, or galaxy morphologies.

There is also evidence that the Tully-Fisher relation and the $D - \sigma$
relation, and other fundamental plane relations, do not necessarily produce
the same results when applied in parallel.  The peculiar velocities derived by
Aaronson \etal (1986) for spirals, and by Lucey \etal (1991) for ellipticals
and S0s, within the cluster A2634, for example, differ significantly.
The effects of the cluster environment upon the two populations may be a very
relevant factor, as is the evidence for an environmental correlation in the
zeropoint of the $D - \sigma$ relation (Lucey \etal 1991; Guzm\'an 1993).
However, the effects of infalling group velocities and, in this case, the
possible inclusion of galaxies from the very nearby cluster A2666 (Scodeggio
\etal 1995) must also be examined.

Our program integrates both optical and \HI observations; we seek to form a
consistent picture of infalling spirals which is sensitive to both gas
depletion and star formation suppression.  Our sample is made up of 329
galaxies, 296 selected from 18 nearby clusters and 33 isolated field galaxies
observed for comparative purposes.  It extends over a wide range of
environments, covering three orders of magnitude in cluster X-ray luminosity and
containing galaxies located throughout the clusters from rich cores out to
sparsely populated outer envelopes.  We have obtained \Halpha rotation curves
to trace the stellar disk kinematics within the potential at high resolution
and to explore the strength of current star formation, \HI line profiles to
map the overall distribution and strength of \HI gas, and \Ib imaging to study
the distribution of light in the underlying, older stellar population.  The
sample contains spirals of all types, and is unbiased by the strength of flux
from \HII regions or by \HI gas detection.  This paper is a companion to Vogt,
Haynes, Herter \& Giovanelli (2004a, \pone), which details the observations
and reduction of the data set, and to Vogt, Haynes, Herter \& Giovanelli
(2004b, \ptwo) which explores the evidence for spiral galaxy infall.  

\section{Evolution in Fundamental Parameters}
\label{sec:FParms}

\subsection{Sampling Bias within the Dynamical Sample} 

\begin{figure*} [hbtp]
  \begin{center}\epsfig{file=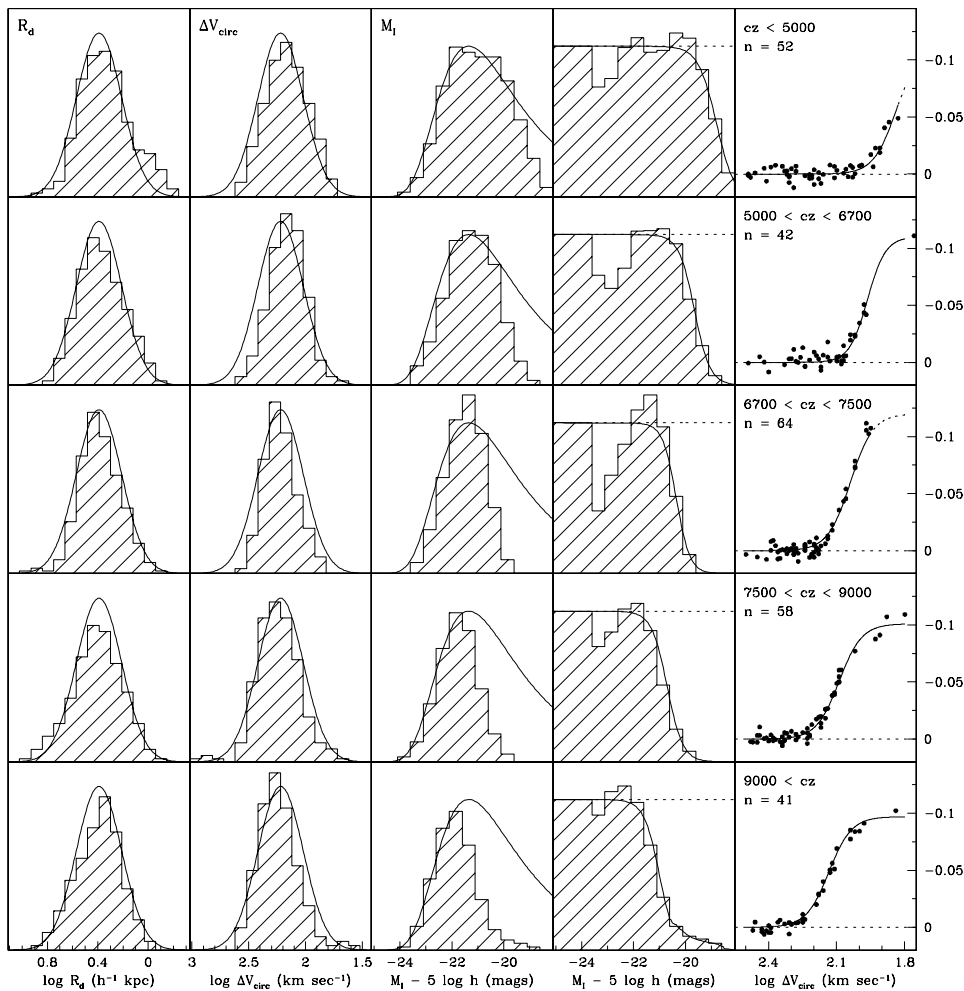,width=5.0truein}\end{center}
  \caption[selection biases diagram]
{Exploration of selection biases throughout the sample, as a function of
redshift.  The first three columns show the distribution of the sample in
size, circular velocity, and luminosity, broken into five redshift bands.  A
Gaussian function is fit to the size and velocity functions, and frozen
across the redshift bins.  A Schechter luminosity function (M$^*$=-22.1,
$\alpha$=-0.50) is similarly applied to the magnitude histogram.  The fourth
column shows the completeness in magnitude varying between 1 (dotted line) and
zero.  The fifth column shows the incompleteness bias in magnitude for each
redshift bin, calculated as discussed in the text, for each galaxy and in a 
smoothed functional form.}
  \label{fig:selRWL}
\end{figure*}

Taken together, these points all serve to motivate our examination of the
fundamental parameters for cluster spirals for environmental variations.  We
proceed to explore the space defined by luminosity, size and circular velocity
through the size-velocity relation, the surface brightness diagram, and the
Tully-Fisher relation.  We first examine three representative variables for
bias.  The process used is derivative of that for our larger Tully-Fisher
program, discussed in great detail within Giovanelli \etal 1997ab.  We begin
by examining the distribution of disk scale length R$_d$, circular velocity
\dV, and \Ib magnitude M$_I - 5$ log h across the dynamical sample.  These
data have been corrected for the effects of extinction, inclination, and
seeing conditions.  We wish to compare the distribution of these fundamental
parameters, which differ in their dependence on distance, against each other.
We will control out the effect of peculiar velocities for each cluster, by
comparison of the standard Tully-Fisher relation (\ie total \Ib magnitude M$_I
- 5$ log h versus terminal circular velocity \dV) offsets with a template
relation derived for a large, overlapping distribution of clusters spread
around the sky (the ``basket of clusters'' approach, where the template
relation has been derived from clusters at cz $\sim$ 6,000 \kms, Giovanelli
\etal 1997).  In order to compute a valid Tully-Fisher relation for each
cluster, we must first evaluate the effects of bias in cluster sampling.

Cluster sampling can be biased by a large number of selection effects,
including the undersampling of faint galaxies, the homogeneous Malmquist bias
(\cf Lynden-Bell \etal 1988) due to the increase in sampled co-moving volume
with distance which scatters distant objects preferentially into the sample,
and variations of galaxy properties with morphological or environmental
variations.  We refer the reader to an extended discussion of the impact of
these factors within Dale \etal 1999b, arguing the negligible impact of all
but the sampling bias towards the bright end of the luminosity function, and
the the discussion ahead in Section~\ref{subsubsec:FPMV} confirming the small
effects of the cluster environment on the Tully-Fisher relation within the
dynamical sample.  The effect of the bias towards bright galaxies, however,
will be to shift the zeropoint of the relation towards brighter magnitudes and
to decrease the apparent slope.

The first three columns of Figure~\ref{fig:selRWL} show the distribution of
disk scale length R$_d$, circular velocity \dV calculated from the terminal
behavior of the \Halpha rotation curves, and total \Ib magnitude M$_I - 5$ log
h across the dynamical sample.  We have chosen to separate the galaxies into
five bins as a function of redshift rather than working on a cluster by
cluster basis, given that the incompleteness is a function of distance, to
increase the number of points within each bin ($\sim$ 50).  There is no strong
trend with R$_d$ or with circular velocity, as expected given that our
selection technique was not tuned to these variables, and we elect to make no
corrections on them.  We do expect a strong trend with absolute magnitude,
given the magnitude dependence of the selection of our survey objects, and
indeed a bias towards increasingly bright intrinsic magnitude shows up clearly
as a function of redshift.

The correction for magnitude is made as follows, in the form of previous
studies (Giovanelli \etal 1997, Dale \etal 1999b).  We select a reference frame
at rest with respect to the CMB, and apply a type-dependent correction to
bring all morphological groups to the same zeropoint ($-0.32$ magnitudes for
Sa and Sab galaxies, $-0.10$ magnitudes for Sb galaxies).  We then model the
completeness in absolute magnitude within each redshift bin as a smoothed step
function varying between 0 and 1, with the form
\begin{equation}
  c(y) = \frac{1}{e^{(y - y_t)/\eta} + 1},
  \label{eq:y_cmplt}
\end{equation}
where $y_t$ is a transitional luminosity, decreasing monotonically with bin
redshift, and $\eta$ controls the rate of decay of the function.  Having fit
$c(y)$ to each bin (see column 4 of Figure~\ref{fig:selRWL}), we can now
evaluate the probable completeness of the sample for each member as a function
of velocity width and redshift.  For each galaxy, we compute a series of trial
magnitudes by assuming the underlying distribution corresponds to the template
form of the Tully-Fisher relation (Giovanelli \etal 1997).  In each trial,
\begin{equation}
  y_{trial} = b_{TF} (\Delta V_{circ} - 2.2) + a_{TF} + f_G~\Delta_x,
  \label{eq:y_trial}
\end{equation}
where $a_{TF} = -21.01$, $b_{TF} = -7.68$, an offset based on the template
scatter $\Delta_x = -0.28 (\Delta V_{circ} - 2.2) + 0.26$, and $f_G$ is drawn
from a Gaussian distribution with zero mean and unit variance.  The value of
$y_{trial}$ is stored if the value of $c(y_{trial})$ for the appropriate
redshift bin is greater than a random number drawn from the range [0,1].  This
process is repeated until 1000 values of $y_{trial}$ have been successfully
stored.  The incompleteness bias for this object is then computed as the
difference between the mean value of $y_{trial}$ and the value expected from
the template relation,
\begin{equation}
  -\Delta  y_{icb} = 0.001 \times \sum_{i = 1}^{1000} y_{trial,i} - 
                     \left \{ b_{TF} (\Delta V_{circ} - 2.2) + a_{TF} \right \}.
  \label{eq:y_icb}
\end{equation}
A smooth step function function $f_{icb}(\Delta V_{circ})$ is then fit to the
distribution of $y_{icb}$ within each redshift bin, as shown in column 5 of
Figure~\ref{fig:selRWL}.  As expected, for a given velocity width the
incompleteness bias in magnitude decreases from zero with redshift.  The
functional form of $f_{icb}(\Delta V_{circ})$ is now subtracted from the
galaxy magnitudes to counteract the tendency to select brighter objects at a
given circular velocity.  The maximum bias term is $-0.1$ magnitudes for our
sample, due to the depth of the cluster sampling and the limited range in
redshift (cz $<$ 13,000 \kms).

\begin{figure*} [htbp]
  \begin{center}\epsfig{file=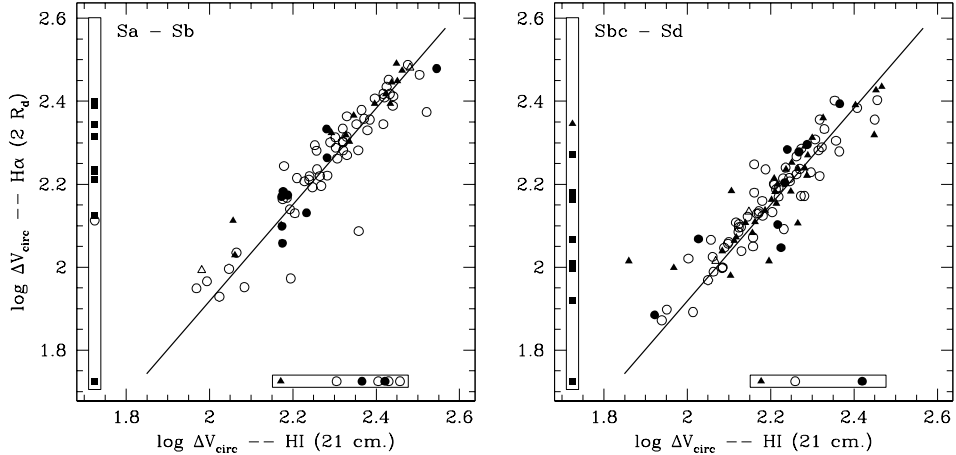,width=5.0truein}\end{center}
  \caption[VV1 diagram]
{Circular velocities of early {\bf (left)} and late {\bf (right)} type spirals
derived from \HI line profiles and from the H$\alpha$ emission line flux at
2R$_d$ for \HI normal field galaxies (filled triangles), the few \HI deficient
field galaxies (open triangles), \HI normal cluster members (open circles),
\HI deficient cluster members (filled circles), and \HI non-detections
(filled squares).  The 27 \HI non-detections are boxed on the left, and the
nine galaxies for which the \Halpha flux does not extend to 2R$_d$ are boxed
below.  We find \dV(\Halpha) = (1.16 $\pm$ 0.03) $\times$
\dV({\rm H{\kern.2em{\smfont I}}}) $-$ (0.049 $\pm$ 0.003).}
  \label{fig:VV1} 
\end{figure*}

The next step in comparing fundamental parameters is to compute the
corrections for the members of each cluster due to peculiar velocities.  Most
of the cluster within the dynamical study were examined within Giovanelli
\etal 1998, and A2151 was examined in Dale \etal 1999b, and the peculiar velocities 
computed therein can be used for this purpose.  We proceed to calculate
peculiar velocity corrections for the remaining clusters A426 and A539, in the
same fashion.  For each cluster, we calculate the mean of the error-weighted
offsets in magnitude from the template relation for the individual galaxies
which are members of the cluster.  This gives us values of 0.150 magnitudes
(-364 \kms) for A426 and 0.070 magnitudes (-277 \kms) for A539.

\subsection{Parameterizing Velocity Profiles}

We begin by examining the velocity profiles of the galaxies.  We have two
independent forms of velocity width measurement: the frequency distribution of
\HI gas, from single-dish \HI line profiles, and the spatial and frequency 
distribution of \Halpha and \fnii line flux. Because the bulk of \HI gas is
found at large radii, the \HI line profile serves as a good measure of
terminal velocity. \Halpha and \fnii rotation curves are similarly useful,
because they typically extend for three to five disk scale lengths, to the
outer portion of the optical disk where the velocity profile becomes fairly
flat.

Optical rotation curves can be characterized by a steeply sloped inner region
(r $\sim \leq$ 1 R$_d$), a transition or elbow point, and then a shallow slope
which gradually tapers off to become flat at large (r $\sim$ 4 R$_d$) radii.
A clear mechanism for creating rising terminal profiles is to truncate the
\Halpha flux before the profile has leveled off.  As discussed in
\ptwo, such a truncation occurs in spiral galaxies interacting with hot 
intracluster gas, and is certainly a function of the cluster environment.  We
have explored previously the good agreement found between velocity widths
derived from the terminal, or near-terminal, behavior of optical rotation
curves and from \HI line profiles (Vogt 1995, Giovanelli \etal 1997a, Dale
\etal 1998).  These measurements depend strongly upon the extent of the
rotation curve, often requiring a minimum extent to the isophote containing
83\% of the \Ib flux (as proposed by Persic \& Salucci 1991, 1995), and
containing a correction term dependent upon the steepness of the terminal
slope of the rotation curve.  This corrective term serves well for peculiar
velocity studies, by allowing calculations of cluster positions to be enhanced
by the inclusion of spiral galaxies well within the cluster cores (typically
\HI deficient and with truncated \Halpha).  However, this assumes that the 
slope-dependent correction term, in effect an extrapolation of the rotation
curve behavior beyond its terminal radii, can be applied universally.

In order to evaluate type-dependent factors we divide our sample into early
(Sa through Sb) and late (Sbc through Sd) types, and allow the zeropoint for
each relation to shift between the two type groups.  We elect to measure the
disk circular velocity at a smaller radius of 2 R$_d$ throughout the dynamical
sample, a distance to which all but nine of the
\Halpha rotation curves extend, and to apply no slope-dependent correction 
term.  Six of the nine have {\it normal} \Halpha extent, being fairly large
galaxies for which R$_d$ $\sim$ 5 \hkpc, and the others are {\it stripped} or
{\it asymmetric}.  We note the work of Willick (1999), who found a reduced
scatter in the Tully-Fisher relationship of the LP10k catalog by measuring
velocity widths at 2.1 R$_d$, in support of the use of V(2R$_d$) in evaluating
mass-to-light ratios.  This width is equal to the \Halpha terminal velocity
width in many cases; lower values are typically found for slow rising profiles
of small or late type spirals which have not reached their full velocity width
at 2 R$_d$.  The most significant difference within the sample in the
percentage rise at 2 R$_d$ lies between early and late type spirals, and we
account for this by examining each morphological group separately and in
combination.

For {\it quenched} galaxies, the \Halpha absorption trough is deep and can be
traced through the nucleus and along the major axis.  It extends to a radius
at or beyond 2 R$_d$ in all but three cases.  This makes it possible to
determine a velocity width from the optical spectrum as is done for emission
line flux, with appropriate corrections between stellar and gas velocities
(see discussion in \pone, also Neistein, Maoz, Rix, \& Tonry, 1999).  In the
few cases where a velocity width can also be measured from the \HI data, it is
found to be in good agreement with that taken from the \Halpha absorption line
flux.

\subsection{Overview of Fundamental Parameter Relations}

We find that the galaxies observed in the general vicinity of the clusters but
more than 3 \hMpc from the center have characteristics similar to those of the
isolated field spirals, and combine the two populations to form a single
population representative of the field.  The region more than 3 \hMpc from the
clusters is typically fairly empty, the exception being that around the A1656
cluster which shows an over-density of spirals at large radii in a bridge
extending towards A1367 (see Figure 1 in \ptwo).  We conducted our analysis
with and without the extended spiral population around A1656, and found no
significant differences.  The three \HI deficient field spirals on our sample
were not used in fits to the field population; we note that they follow the
characteristics of the \HI normal field population more than the
\HI deficient cluster population.

We examined the three relations by dividing our sample galaxies into various
subsets.  All galaxies within 3 \hMpc of the cluster cores which were not
background or foreground superpositions were treated as cluster members,
including both accreted spirals and infalling subgroups.  We inspected each
individual cluster in turn, divided the clusters into X-ray hot and X-ray cold
groups, and binned galaxies according to their individual \HI deficiency.  The
four clusters with X-ray temperatures greater than 4 keV (A1656, A426, A2199 and
A1367) made up the X-ray hot group, and the other clusters and groups were all
considered to be X-ray cold.  

We found that the offsets from the \HI normal field galaxies within the
sampled clusters correlated most strongly with \HI deficiency, and cluster X-ray
temperature.  There is, of course, a strong overlap between the \HI deficiency
and X-ray hot cluster membership (\cf Magri \etal 1988).  \HI deficient members
of X-ray hot clusters showed strong trends and \HI normal members of X-ray cold
clusters showed the least deviation from the field spiral population,
unsurprisingly, but we note that \HI normal members of X-ray hot clusters show
more (at marginal significance, however) of a deviation from the field
population than do \HI deficient members of X-ray cold clusters.  The most
deviant galaxies within the cluster sample were late type members of X-ray hot
clusters, so gas deficient that no generally no \HI gas had been detected (to
limits of $\sim$ 5 $\times$ 10$^8$ M$_{\odot}$).  Figures \ref{fig:FPRV}
through \ref{fig:FPMV} illustrate these trends, as discussed below.

\subsubsection{The Size-Velocity Relation}

\begin{figure*} [htbp]
  \begin{center}\epsfig{file=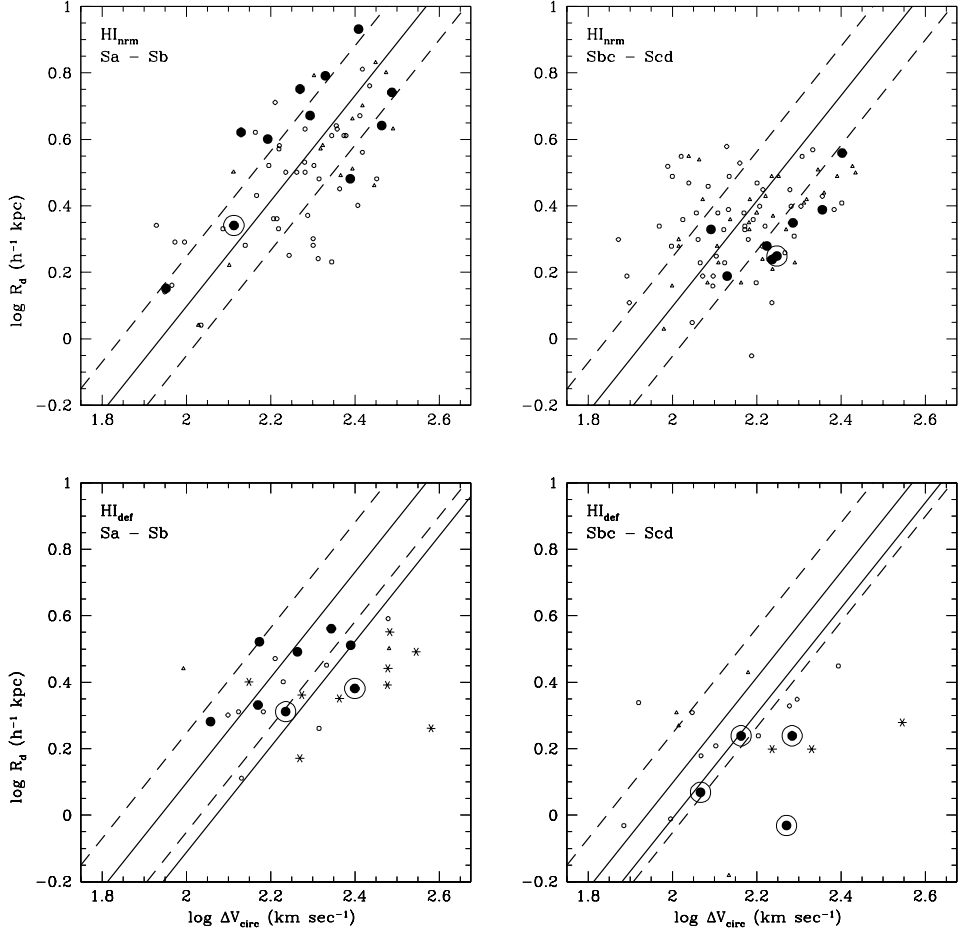,width=5.0truein}\end{center}
  \caption[FP-RV diagram]
{Exponential disk scale length R$_d$ as a function of circular velocity \dV,
divided into early {\bf (left)} and late {\bf (right)} morphological types,
and \HI normal {\bf (top)} and \HI deficient {\bf (bottom)} spirals.  Small
open triangles represent isolated field galaxies, and small open circles
galaxies within cool clusters or located more than 900 \hkpc from the core of
a warm (kT $>$ 4 keV) cluster.  Core members of warm clusters are shown with a
large solid circle; those few within the cores of the three hottest clusters
(A1656, A426, and A2199) are encircled, and {\it quenched} galaxies are marked
with asterisks.  Early and late types were offset by $\pm 0.081$ in log R$_d$
to bring the zeropoints of the \HI normal morphological groups into agreement,
and an error weighted fit to the complete \HI normal population, with dashed
1$\sigma$ limits, is shown in all four windows.  The second solid line on the
lower plots shows the $\sim$ 30\% offset in R$_d$ to the \HI deficient
galaxies for each morphological group.}
  \label{fig:FPRV}
\end{figure*}

Figure \ref{fig:FPRV} shows the relation between disk scale length and
circular velocity throughout the dynamical sample.  (Disks and bulge surface
brightness profiles were deconvolved, and both disk scale lengths and
velocities were corrected for inclination, as discussed in \pone.)  We present
an average of $y-$on$-x$ and $x-$on$-y$ error weighted, 3$\sigma$-clipped
least-squares fits to the 159 galaxy \HI normal population, with the centered
variables $x$ = log \dV $-$ 2.20 and $y$ = log R$_d$ $-$ 0.38, and find the
relation log R$_d$ = (1.59 $\pm$ 0.18) $\times$ log \dV $-$ ($3.15 \pm 0.01$)
with early/late morphological types lying at $\mp 0.081$.  (The offset between
field and cluster galaxies is negligible for both \HI normal type groups.)

We find an offset from the \HI normal early type population to the 28 early
type \HI deficient galaxies of $-0.21 \pm 0.15$ in log R$_d$.  The nine {\it
quenched} galaxies form the extreme edge of this population, falling at $-0.35
\pm 0.18$ in log R$_d$, well below the \HI normal sample.  There are
substantially fewer \HI deficient late type galaxies, and they exhibit more
scatter in all three of the planes which we will examine in
Figures~\ref{fig:FPRV} through~\ref{fig:FPMV}.  The 21 detected late type \HI
deficient galaxies lie at a similar offset of from the late type \HI normal
spirals, at $-0.13 \pm 0.21$, and the three late type {\it quenched} spirals
fall at $-0.46 \pm 0.16$.

One advantage of examining the size-velocity relation is that it is
independent of the direct effects of changes in surface brightness, and allows
us to focus on the more static properties of disk size and velocity (strongly
tied to mass).  If the \Ib disk scale length is unaffected by the suppression
of star formation in the disk, as predicted, for example, for models of tidal
interactions (Gnedin 2003), this implies that the quenched spiral disks formed
at a redshift $z \sim$ 1.  We treat this epoch as an upper limit as we assume
(1) that the disk scale lengths are unlikely to have been larger in the past
(\ie stars were built up along the disk from the inside-out), (2) that halo or
disk truncation would have only shortened the scale lengths), and (3) that
galaxies which are still identifiable as spirals are unlikely to have
undergone a major merger, and as we are working at near-infrared wavelengths
(less affected by the instantaneous star formation rate than \Bb).  This is in
reasonable agreement with observations of disks at high redshift (Lilly \etal
1998, Vogt 2000, Le F\`evre \etal 2000).  Though the shift of \foii3727\AA
~out of the optical passband makes spectral redshift determinations for star
forming galaxies above a redshift $z \sim$ 1.2 fairly difficult, deeply imaged
fields such as the two Hubble Deep Fields suggest that the virialized field
disk population drops off between $z \sim$ 1 and 1.5 (Vogt 2000).

\subsubsection{The Surface Brightness Diagram}

\begin{figure*} [htbp]
  \begin{center}\epsfig{file=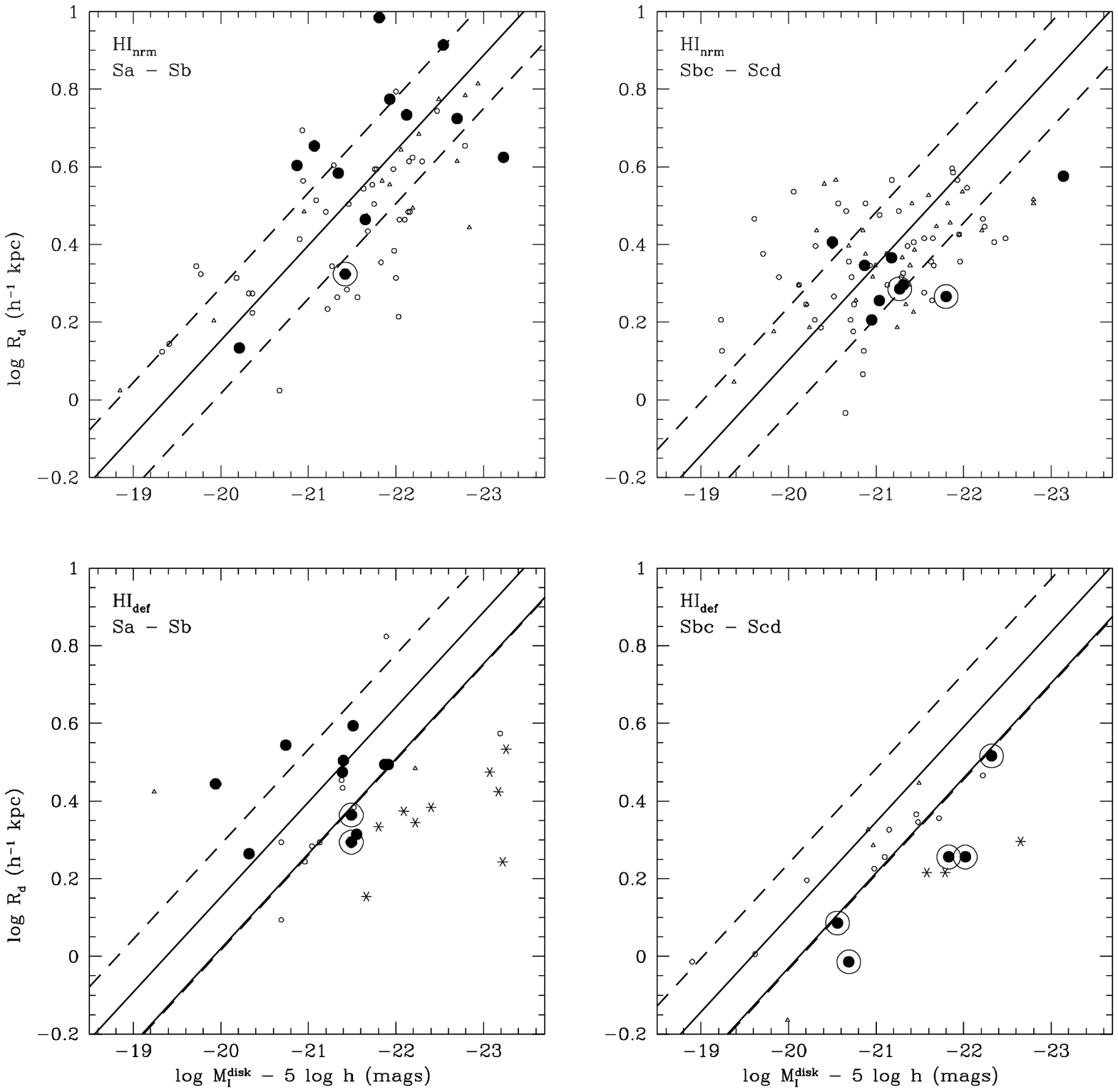,width=5.0truein}\end{center}
  \caption[FP-RM diagram]
{Exponential disk scale length R$_d$ as a function of disk \Ib luminosity
M$_I^{disk} - 5$ log h, divided into early {\bf (left)} and late {\bf (right)}
morphological types, and \HI normal {\bf (top)} and \HI deficient {\bf
(bottom)} spirals.  Symbols are as in Figure \ref{fig:FPRV}.  Early and late
types were offset by $\pm 0.064$ in log R$_d$ to bring the zeropoints of the
\HI normal morphological groups into agreement, and the {\it quenched}
population was brightened by 0.70 magnitudes.  An error weighted fit to the
complete \HI normal population, with dashed 1$\sigma$ limits, is shown in all
four windows.  The second solid line on the lower plots shows the $\sim$ 25\%
offset to the \HI deficient galaxies for each morphological group.}
  \label{fig:FPRM}
\end{figure*}

Figure \ref{fig:FPRM} shows the surface brightness properties of the sample.
The fit to the \HI normal galaxies is log R$_d$ = ($-0.25 \pm$ 0.03) $\times$
(M$_I^{disk} - 5$ log h) $-$ (4.77 $\pm$ 0.01), with the fit to the centered
variables $x$ = M$_I^{disk} - 5$ log h $+$ 21.2 and $y$ = log R$_d$ $-$ 0.38
and early/late types lying at $\mp 0.064$.  The \HI normal cluster galaxies
again follow the distribution of the field spirals.  We have converted de
Jong's (1995) \Bb surface brightness relation for a range of field spirals
into \Ib, for comparative purposes.  Assuming an average \BI = 1.87, it
becomes log R$_d$ = $-0.20$ $\times$ (M$_I^{disk} - 5$ log h) $-$ 3.79, in
reasonable agreement.

The {\it quenched} population is expected to have faded substantially in \Ib,
given its evolutionary track.  We apply a brightening factor of 0.70
magnitudes, (1) in line with projected fading from stellar populations, (2) to
match the offset in R$_d$ in the R$-$V relation with that remaining in surface
brightness, and (3) to balance the offset of the {\it quenched} galaxies
within the M$-$V relation, after examining all three fundamental parameter
relations.  The offset for early type \HI deficient galaxies is then $-0.14
\pm 0.16$ in log R$_d$, and the {\it quenched} spirals are again additionally
offset, at $-0.40 \pm 0.10$.  The late type \HI deficient galaxies are offset
by $-0.13 \pm 0.11$ in R$_d$, as in the R$-$V plane, and we place the late
type {\it quenched} galaxies at $-0.35 \pm 0.07$ in R$_d$.  We note that the
blue {\it asymmetric} galaxies are offset 0.50 mag brighter than the red,
which supports their placement in a starbursting phase.

\subsubsection{The Tully-Fisher Relation}
\label{subsubsec:FPMV}

\begin{figure*} [htbp]
  \begin{center}\epsfig{file=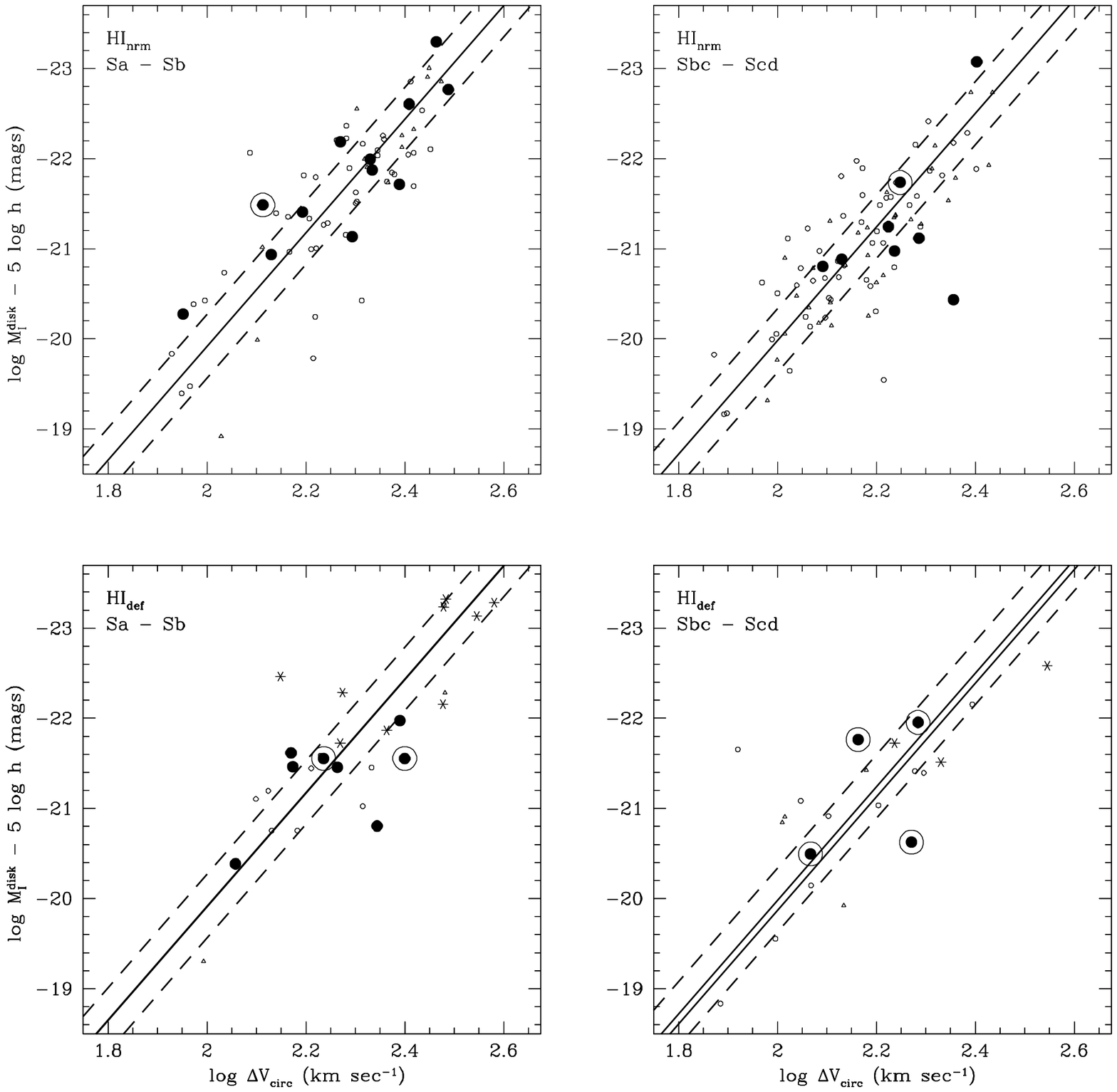,width=5.0truein}\end{center}
  \caption[FP-MV diagram]
{Disk \Ib luminosity M$_I^{disk} - 5$ log h as a function of circular velocity
\dV, divided into early {\bf (left)} and late {\bf (right)} morphological
types, and \HI normal {\bf (top)} and \HI deficient {\bf (bottom)} spirals.
Symbols are as in Figure \ref{fig:FPRV}.  Early and late types were offset by
$\mp 0.065$ in M$_I^{disk} - 5$ log h to bring the zeropoints of the \HI
normal morphological groups into agreement, and the {\it quenched} population
was brightened by 0.70 magnitudes.  An error weighted fit to the complete \HI
normal population, with dashed 1$\sigma$ limits, is shown in all four windows.
The second solid line on the lower plots shows the minimal offset to the
\HI deficient galaxies for each morphological group.}
  \label{fig:FPMV}
\end{figure*}

Figure \ref{fig:FPMV} shows the Tully-Fisher relation for our data set.  We
find an \HI normal spiral relation in agreement with the literature, in the
range of other \Ib results (\cf Pierce \& Tully 1992, Giovanelli \etal 1997b,
Verheijen 2001).  Recall that we are using disk rather than total galaxy
magnitudes, with mean bulge fractions are (13 $\pm$ 9)\%, and velocity widths
measured at 2 R$_d$ as our cardinal variables, and so corrections must be made
to transfer from one survey to another to compare results.  An average of
$y-$on$-x$ and $x-$on$-y$, error weighted least-squares fits to the centered
variables $x$ = \dV $-$ 2.20 and $y$ = M$_I^{disk} - 5$ log h $+$ 21.2 yields
the relation M$_I^{disk} - 5$ log h = (-6.31 $\pm$ 0.33) $\times$ log \dV $-$
(7.26 $\pm$ 0.01), with early/late types lying at $\pm 0.065$.

The offset for early type \HI deficient galaxies is $0.003 \pm 0.42$ in log
M$_I^{disk} - 5$ log h (shifting the zeropoint to brighter magnitudes), while
the {\it quenched} spirals are offset by $-0.10 \pm 0.56$.  The offsets are
similarly small for late type spirals, ($0.06 \pm 0.47$, and $0.29 \pm 0.46$
for the three late type {\it quenched} spirals).  The alignment of the three
sub-populations, the \HI normal galaxies, and \HI deficient, and the {\it
quenched} spirals, in each morphological bin, supports the proposed
brightening correction of 0.70 magnitudes for the {\it quenched} population.
If the disk scale lengths of \HI deficient galaxies are decremented by $\sim$
25\% as suggested by their distribution in the R$-$V and the R$-$M planes,
then we may in fact be measuring circular velocities at 1.5 R$_d$ rather than
at 2.0 R$_d$.  When we refit these velocity profiles at 2.5 R$_d$, the offset
from the \HI normal relation for those that extended that far in \Halpha
became insignificant.  As the galaxies for which the extent of \Halpha is most
truncated (between 2.0 and 2.5 R$_d$) may be those most altered by the
stripping process, we re-integrated them into the revised fit by comparing the
velocities measured at 2 R$_d$ for the \HI deficient population with
velocities measured at 1.5 R$_d$ for the \HI normal population; again the
already small offset became negligible.

In contrast to the R$-$V and R$-$M planes, we find that the \HI deficient
galaxies show no significant offset relative to the \HI normal population,
with the exception of the {\it quenched} galaxies.  This suggests that,
although cluster spirals are smaller than their field counterparts, any
alterations to the global brightness or mass either preserve the mass-to-light
ratio (moving along the Tully-Fisher relation rather than away from it), or
take effect significantly more slowly than the fast ($\Delta$t $\sim$ 10$^7$
years) \HI stripping processes.

Early work with \Halpha optical rotation curves (Rubin, Whitmore, \& Ford
1988; Whitmore, Forbes, \& Rubin 1988; Forbes \& Whitmore 1989) suggested that
the observed velocity profiles declined at large optical radii more in cluster
spirals than in the field, implying the presence of substantially stripped or
stunted halos, but this result has not been supported by later observations
(Distefano \etal 1990; Amram \etal 1993 two-dimensional velocity maps;
Sperandio \etal 1995; Vogt 1995; Dale \etal 1999a).  We have examined the
inner and outer slopes of the optical rotation curves calculated by a range of
methods, and determined stellar mass-to-light ratios for our sample calculated
by a velocity decomposition of the bulge, disk and halo, and find no evidence
for significant offsets as a function of \HI deficiency or cluster X-ray gas
temperature there either.  (Note that apparently larger outer slopes of
cluster spirals can be accounted for by the radial truncation of the \Halpha
emission, leading to a measurement biased towards the rising inner part of the
curve.)  Unfortunately, the \Halpha velocity profiles of cluster (and of many
field) galaxies do not extend to large enough radii for us to constrain the
halo core radius or shape significantly.  We have performed bulge, disk, and
halo fits to the combination of velocity profiles and deconvolved surface
brightness profiles (Vogt 1995), but we are unable to provide direct evidence
for or against halo truncation within the optical radius.

One could argue, given the lack of change to the full velocity profiles, that
the three-plane data shown in Figures~\ref{fig:FPRV} through \ref{fig:FPMV}
support the case for dark matter domination over baryons in the inner regions
of spirals disks.  However, we must contrast this with the work of Debattista
\& Sellwood (2000), who argue that a dominant inner halo would drag down the
rotational speed of bars in barred galaxies, and of Dalcanton and of van den
Bosch (2000), whose rotation curves of late type, low surface brightness
spirals cannot be fit in the cores with any of the standard halo models (\eg
Hernquist, NFW).

The quenched spirals are found to lie 0.70 magnitudes fainter than the \HI
normal relation, where one would expect a spiral to lie if star formation had
previously been halted by removing the gas reservoir without significantly
disturbing the stellar structure of the disk (\cf Kannappan, Fabricant, \&
Franx 2002).  This is particularly interesting to note, in view of recent work
on a Tully-Fisher relation by Hinz, Rix, \& Bernstein (2001) and Neistein,
Maoz, Rix, \& Tonry (1999), finding offsets $\leq 0.2$ magnitudes from late
type spirals for 18 Coma S0s.  In contrast, Magorrian (2000) has used
constant-anisotropy spherical dynamical models to study 18 ellipticals and
derived an offset $\sim$ 1 magnitude fainter than normal spirals for them.
Clearly, morphological type classification (and inclination angles) play an
important role in interpreting these results.

In summary, the simplest way to reconcile the offsets shown within the three
planes of Figures \ref{fig:FPRV} through \ref{fig:FPMV} is through a decrease
in disk scale length of $\sim$ $25$\% for \HI deficient galaxies, with an
additional fading of 0.70 magnitudes for the {\it quenched} spirals.  The {\it
quenched} spirals represent the extreme edge of the truncated population, and
the \HI normal members of hot cluster cores, though small in number, also
follow this trend.

\section{Star Formation Properties}
\label{subsec:SFProps}

Star formation rates have been studied previously in clusters by use of
broadband colors, narrow-band \Halpha photometry, and \Halpha equivalent
widths.  Kennicutt \etal (1984) observed that many cluster spirals had star
formation properties equivalent to those of field spirals and that \HI
deficiency did not necessarily imply a change in the overall star formation
rate.  Recent work on circumnuclear starbursts has identified probable
tidal-interaction candidates in the cores of rich clusters (Moss \& Whittle
2000; see also Andersen 1999) and extensive analyses of the Virgo cluster
(Koopmann 1997; Gavazzi \etal 2002a, 2002b) has targeted star formation 
suppression in the outer parts of cluster disks through a variety of \HI 
gas stripping and redistribution mechanisms.

We looked for evidence within the dynamical sample supporting Koopmann's
finding that a substantial fraction of cluster spirals are assigned early
spiral type Hubble types due to low global star formation rates, though their
central light concentrations are more representative of late type spirals (\ie
stripped late type spirals are systematically misclassified as early types).
Lacking central concentration measurements, we examined \Ib bulge-to-total
(B/T) fractions as a function of galaxy type, but found no significant trend
with \HI deficiency.  We do not find evidence for \HI deficient, late type
interlopers, with small bulge fractions, within the early type
classifications.  However, we do note that the {\it asymmetric} late type
galaxies have larger B/T fractions and redder \BI colors than {\it normal}
late type spirals, suggesting a shift toward certain early type qualities.
There is an offset of $\pm 0.075$ in log R$_d$ between late and early
morphological type galaxies regardless of \HI gas content
(Figures~\ref{fig:FPRV} and \ref{fig:FPRM}).  Because of this, stripped late
type galaxies classified as early types could contribute to the smaller offset
observed for \HI deficient early type galaxies than for late types, as they
would fall closer to the \HI normal early type galaxy relation than to that
for late types.

We lack \Halpha photometry and multi-band aperture colors for our sample, but
the distribution and extent of \Halpha along the 1\arcsec slits we placed
along the galaxy major axes provides a tracer of young star formation.  Figure
10 in \ptwo\ shows the distribution of \Hax emission throughout the dynamical
sample.  \Hax is shown as a function of R$_d$, thus taking into account the
decrement of 25\% in R$_d$ found for the \HI deficient galaxies.  An average
of $y-$on$-x$ and $x-$on$-y$, error weighted least-squares fits, conducted on
the centered variables $x =$ log R$_d$ $-$ 0.38 and $y =$ log \Hax $-$ 0.90,
yields the one-to-one relation log~\Hax = (0.85 $\pm$ 0.06) $\times$ log R$_d$
+ (0.62 $\pm$ 0.09) for \HI normal galaxies, with no significant offset ($\mp
0.01$ in log \Hax) between morphological types.  The early type \HI deficient
spirals have truncated emission for their size, by $-0.13$ $\pm$ 0.07 in log
\Hax (equivalent to a decrease of 26\% in \Hax).  In contrast, the \Halpha
emission in the \HI deficient late type spirals shows no significant change in
length ($-0.01$ $\pm$ 0.09).  If \Hax was unchanged by the stripping process
and by the cause of the change in R$_d$, we would expect an apparent offset of
$+0.12$, due to the decrement of 25\% in log R$_d$ in \HI deficient galaxies.
The observed decrement of R$_d$, isolated in the plots of fundamental
parameters for spirals of {\it all morphological types}, is not echoed in
\Halpha truncation, which suggests that the change in R$_d$ is not an artifact
of the \HI stripping process.  The extent of the \Halpha absorption trough
along the major axes of the {\it quenched} spirals is significantly more
truncated than the distribution of the \Halpha emission line for \HI deficient
galaxies.  The distribution of the old stellar population contributing to the
\Hax absorption of the {\it quenched} spirals may be partly responsible for
this extreme truncation, if disks are built up from the inside to the outside
over time.

\begin{figure*} [htbp]
  \begin{center}\epsfig{file=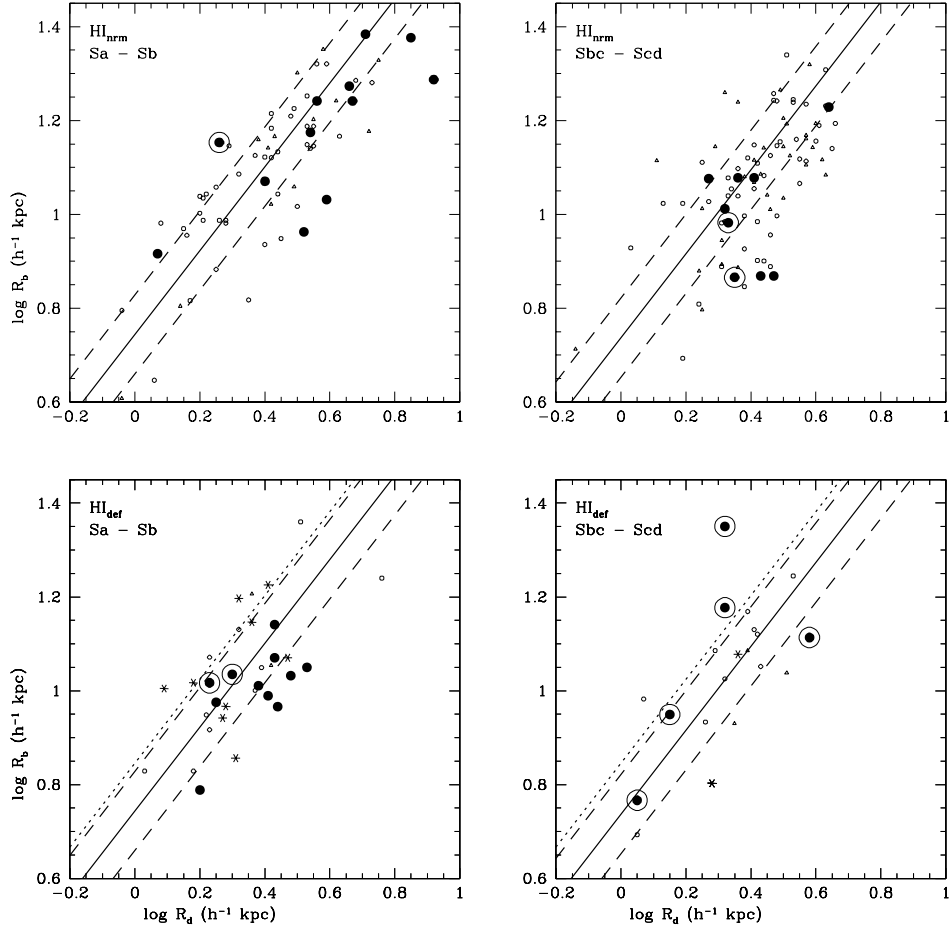,width=5.0truein}\end{center}
  \caption[Rd-Rb plot]
{\Bb radius R$_b$ as a function of exponential disk scale length R$_d$,
divided into early {\bf (left)} and late {\bf (right)} types, and \HI normal
{\bf (top)} and \HI deficient {\bf (bottom)} spirals.  Symbols are as in
Figure \ref{fig:FPRV}.  An error weighted fit to the \HI normal population is
shown on the top two plots, with 1$\sigma$ limits, and reproduced on the
bottom two plots.  The second line (dotted) on the lower plots designates the
predicted relation {\it if} R$_d$ was decremented by 25\% for \HI deficient
galaxies, but R$_b$ remained unchanged; the distribution of early type \HI
deficient galaxies is in clear disagreement.}
  \label{fig:RdRb}
\end{figure*}

A contrasting result from a recent large-scale study of 510 rotation curves
(Dale \etal 2001) found weaker trends in the extent and asymmetry of the
\Halpha flux with clustercentric radius.  The strong selection bias of the
survey towards galaxies with strong, extended \Halpha or \fnii 6584\AA\
emission, chosen by eye to lack a dominant bulge component (Dale \etal 1997),
and of late type (70\% are Sbc or later, Sc galaxies alone make up 52\%, Dale
\etal 1999a) in effect has biased the sample against the galaxies which show
the most change in \Halpha, and explains our contrasting results.

We note that the combined effects of disk size decrement and \Hax truncation
can also well explain the apparent increase in rotation curve outer slopes for
certain cluster spirals reported in Vogt (1995).  By truncating the \Halpha
flux within the inner regions of the rotation curve, before the curve has
leveled off to a terminal straight relation, and decrementing R$_d$, one can
measure an inflated, high slope by fitting the {\it decremented} final extent
of the rotation curve at smaller radii.

If the changes in R$_d$ are due to the galaxies within in the rich cluster
cores having formed at an early epoch, the truncation of \Hax is best
attributed to the direct effects of the galaxy local environment, be it \HI
gas stripping through interaction with the intracluster medium, galaxy
harassment or tidal effects.  This is supported by the distribution of the
{\it asymmetric} spirals through the plot.  The \HI deficient {\it asymmetric}
spirals follow the relation of the general \HI deficient spirals.  Those which
are still fairly gas-rich, however, and thus assumed have only recently begun
the stripping process, show an interesting feature.  When \Hax is defined as
the {\it shorter} extent of \Halpha along either side of the disk, the
gas-rich asymmetric spirals follow the relation of the general \HI deficient
spirals.  When we examine the {\it longer} extent of \Halpha, however, we find
that these galaxies follow the relation of \HI normal spirals.  This implies,
in support of Koopmann 1997, that the complete stripping process is
systematically truncating star formation in the outer parts of the disk.

Figure~\ref{fig:RdRb} shows similarly that the \Bb radii R$_b$ are smaller in
the \HI deficient early type galaxies than in the \HI normal sample; as with
the \Halpha extent, we do not see an equivalent drop in the late type
galaxies.  We find log~R$_b$ = (0.89 $\pm$ 0.06) $\times$ log R$_d$ +
(0.71e$\pm$ 0.08) for \HI normal galaxies, with no significant offset ($\mp
0.01$ in log R$_b$) between morphological types.  The early type \HI deficient
cluster spirals are offset from the \HI normal field population by 0.02 $\pm$
0.09 in log R$_b$.  (There is a small offset to the {\it quenched} population,
at 0.05 $\pm$ 0.10.) If R$_b$ was unchanged by the stripping process, we would
expect an offset of 0.11, caused by the decrement of 25\% in log R$_d$ in \HI
deficient galaxies.  The late type \HI deficient cluster spirals are offset
from the \HI normal population by 0.08 $\pm$ 0.10, suggesting that R$_b$ is
unchanged by the stripping process for recognizably late type spirals.  The
scatter in the R$_b$ $-$ R$_d$ relationship is large, but this suggests that
R$_b$, like \Hax, is decremented in the stripping process, while the decrement
in the R$_d$ may be a remnant of an early formation epoch.  Assuming a
constant young star formation rate, it is possible that a formation-era
initial decrement in R$_b$ could be subsumed within later star formation and
radially increasing disk growth.

\begin{figure*} [htbp]
  \begin{center}\epsfig{file=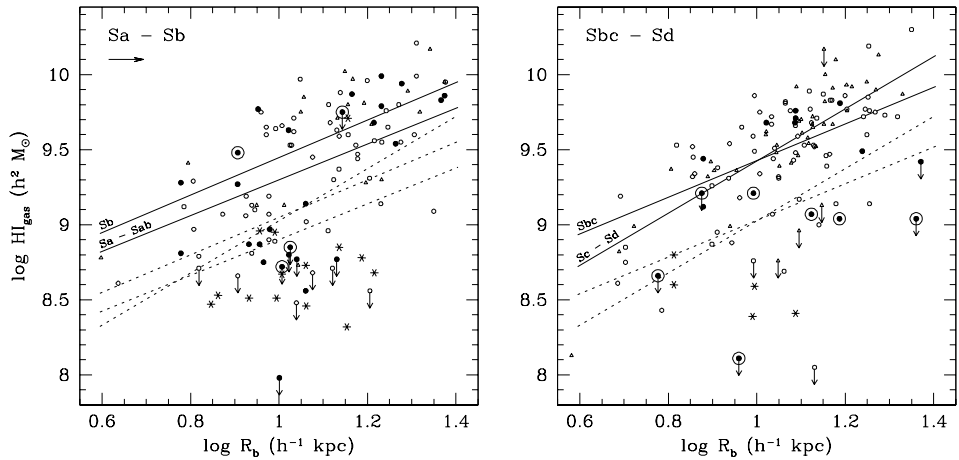,width=5.0truein}\end{center}
  \caption[HI-Rb plot]
{The total \HI gas mass vs. blue radius R$_b$ for early ({\bf left}) and late
type ({\bf right}) galaxies.  Symbols are as in Figure \ref{fig:FPRV}. Upper
limits are drawn for all undetected galaxies, with the exception of the {\it
quenched} population (hidden there to simplify the plots, as all but two are
undetected).  The solid lines show the normal relations used to calculate \HI
deficiency as a function of R$_b$ for various galaxy types, and the dashed
lines below the \HI gas upper limit for the \HI deficient population (less
than 40\% of the expected gas mass) for each type.  The relations for Sb
(left) and Sbc (right) fall virtually on top of each other; the relation for
earlier type (left) galaxies lies at 70\% of the Sb line, while the relation
for later types has a much stronger dependence on R$_b$.  The \HI deficiency
limit for late (Sbc through Scd) types has been reproduced on the lefthand
plot.  The arrow in the upper lefthand corner shows the decrement in R$_b$
that the early type \HI deficient galaxies may have endured in the process of
being stripped, which could lead to an underestimate of 20\% in the \HI gas
loss.}
  \label{fig:HIRb}
\end{figure*}

Figure~\ref{fig:HIRb} shows the distribution of \HI gas content across the
dynamic sample as a function of R$_b$ and morphological type, the two
variables commonly used in modeling expected gas masses (\cf Solanes,
Giovanelli, \& Haynes 1996 and references therein).  The distribution of \HI
mass is fairly continuous; there is no zone of avoidance between normal and
deficient states.  Almost none of the galaxies for which we have an upper
limit on the gas content lie above the zone of substantial (40\%) \HI
deficiency.  The difference in the predicted amount of \HI gas as a function
of morphological type is far less than the distinction between the \HI normal
and \HI deficient categories.  However, due to the steep slope of the relation
for late type spirals, if any of the \HI deficient galaxies classified as
early type are late type spirals which have undergone a partial morphological
transformation, the degree of \HI gas mass loss could be under/over-estimated
by up to a factor of two for large/small values of R$_d$.  We note also that
the decrement in R$_b$ observed for early type, \HI deficient galaxies
(Figure~\ref{fig:RdRb}) would translate into an underestimate of \HI gas
deficiency of 20\% for these galaxies.

\section{Conclusions}

We have conducted a study of optical and \HI properties of spiral galaxies to
explore the role of gas stripping as a driver of morphological evolution in
clusters.  For \HI deficient local cluster spirals (with less than 40\% of the
predicted atomic gas mass), we observe a decrease of a factor of $25$\% in \Ib
disk scale lengths relative to \HI normal spirals in spirals of all
morphological types.  This may be a relic of early galaxy formation, caused by
the disk coalescing out of a smaller, denser halo (\eg higher concentration
index) or reflect an environmental effect (\eg truncation of the hot gas
reservoir at large radius due to the high local density of neighboring
galaxies), or it may be be the product of post disk-formation effects (\eg gas
stripping, tidal interactions).  The few \HI normal spirals observed in the
cores of rich clusters are similarly decremented, in support of the former.

\Bb radii R$_b$ are decreased by 25\% relative to \HI normal field
galaxies in \HI deficient early type spirals (Sa through Sb).  They show an
additional decrease of 55\% in the extent of the \Halpha flux along the disk,
on or within the stripping radius predicted for the \HI gas loss from models.
Gas-rich {\it asymmetric} spirals reflect this trend on the truncated side of
the disk, implying that it is caused by the gas stripping process.  Late type
spirals (Sbc through Scd) exhibit neither trend, suggesting that either gas
removal is less effective within the potential or that once significant star
formation suppression occurs these galaxies are no longer identified as late
type spirals.

For star forming spiral galaxies associated with all of the clusters, we find
no significant trend in stellar mass-to-light ratios or circular velocities
with \HI gas content, morphological type, or clustercentric radius.  This
could be interpreted as support for dark matter domination over baryons well
within the optical radius of disks.

In summary, we have explored the formation and evolution of spiral galaxies in
local clusters through a combination of optical and \HI properties.  We find
evidence that the spirals within rich cluster cores, believed to coalesce from
their halos at an early epoch, formed with intrinsically smaller disks than
the local field population.  We have explored the relationship between
\HI gas stripping and the consequential suppression of young star formation in
spiral galaxies.  We find that the ram pressure stripping and the
suppression of star formation both occur quickly within spirals
infalling into an intracluster medium of hot gas, but find little
evidence for substantial mass stripping beyond that of the \HI gas.

\section{Acknowledgments}

The data presented in this paper are based upon observations carried out at
the Arecibo Observatory, which is part of the National Astronomy and
Ionosphere Center (NAIC), at Green Bank, which is part of the National Radio
Astronomy Observatory (NRAO), at the Kitt Peak National Observatory (KPNO),
the Palomar Observatory (PO), and the Michigan--Dartmouth--MIT Observatory
(MDM). NAIC is operated by Cornell University, NRAO by Associated
Universities, inc., KPNO and CTIO by Associated Universities for Research in
Astronomy, all under cooperative agreements with the National Science
Foundation. The MDM Observatory is jointly operated by the University of
Michigan, Dartmouth College and the Massachusetts Institute of Technology on
Kitt Peak mountain, Arizona. The Hale telescope at the PO is operated by the
California Institute of Technology under a cooperative agreement with Cornell
University and the Jet Propulsion Laboratory.  We thank the staff members at
these observatories who so tirelessly dedicated their time to insure the
success of these observations.
We also thank Sc project team members John Salzer, Gary Wegner, Wolfram
Freudling, Luiz da Costa, and Pierre Chamaraux, and also Shoko Sakai and Marco
Scodeggio, for sharing their data in advance of publication.  
The text of this manuscript was much improved by careful reading on the part
of Richard Ellis.  We thank the anonymous referee for helpful comments on the 
manuscript.

N.P.V. is a Guest User, Canadian Astronomy Data Center, which is operated by
the Dominion Astrophysical Observatory for the National Research Council of
Canada's Herzberg Institute of Astrophysics.  This research has made use of
the NASA/IPAC Extragalactic Database (NED) which is operated by the Jet
Propulsion Laboratory, California Institute of Technology, under contract with
NASA, and NASA's Astrophysics Data System Abstract Service (ADS).  This
research was supported by NSF grants AST92--18038 and AST95--28860 to
M.P.H. and T.H., AST90--23450 to M.P.H., AST94--20505 to R.G., and
NSF--0123690 via the ADVANCE Institutional Transformation Program at NMSU, and
NASA grants GO-07883.01-96A to N.P.V. and NAS5--1661 to the WFPC1 IDT.
N.P.V. acknowledges the generous support of an Institute of Astronomy rolling
grant from PPARC, reference number PPA/G/O/1997/00793.


\end{document}